\documentclass{cta-author}

{}
{}
{}

\usepackage{stackengine}
\usepackage{booktabs}
\usepackage{dcolumn}
\usepackage{array}
\usepackage{multirow}
\usepackage{graphicx}
\usepackage[caption=false,font=footnotesize]{subfig}
\usepackage{fixltx2e}
\usepackage{upgreek,bm}

\newcommand*{\Scale}[2][4]{\scalebox{#1}{$#2$}}%
\begin{document}

\title{Sidelobe Level Reduction in ACF of NLFM Waveform}

\author{\au{Roohollah Ghavamirad$^{\;1}$, }
\au{Mohammad Ali Sebt$^{\;1*}$}
}

\address{\add{1}{Department of Electrical Engineering K. N. Toosi University of Technology, Tehran, Iran}
\email{sebt@kntu.ac.ir}}

\begin{abstract}
In this paper, an iterative method is proposed for nonlinear frequency modulation (NLFM) waveform design based on a constrained optimization problem using Lagrangian method. To date, NLFM waveform design methods have been performed based on the stationary phase concept which we have already used it in a previous work. The proposed method has been implemented for six windows of Raised-Cosine, Taylor, Chebyshev, Gaussian, Poisson, and Kaiser. The results reveals that the peak sidelobe level of autocorrelation function reduces about an average of 5 dB in our proposed method compared with the stationary phase method, and an optimum peak sidelobe level is achieved. The minimum error of the proposed method decreases in each iteration which is demonstrated using mathematical relations and simulation. The trend decrement of minimum error guarantees convergence of the proposed method.
\end{abstract}

\maketitle
\section{Introduction}
Using a short simple pulse for sending in a radar system needs to employ a high power transmitter which is expensive and easy to intercept. If the pulse width increases, the required range resolution to detect the several targets is severely degraded; therefore, the best solution is to modulate the pulse \cite{1,2}. By performing modulation also known as pulse compression, bandwidth and signal-to-noise ratio (SNR) are increased and range resolution improves \cite{3,4,5,6,7,8}, but the existence of high sidelobe level in the autocorrelation function (ACF) is annoying \cite{9,10}. Pulse compression is done using various methods such as phase coding (Barker code), amplitude weighing, linear frequency modulation (LFM), and NLFM \cite{11,12,13}.

In the phase coding and amplitude weighing methods, due to the phase discontinuity and variable amplitude, the mismatch loss increases in the receiver \cite{14}; therefore, the use of LFM signals increases significantly due to continuous phase and constant amplitude. Although, the LFM method shows clear advantages over the phase coding and amplitude weighing methods, but the high sidelobe level is still annoying because of masking the smaller targets by sidelobes of bigger targets; thus, the NLFM method is used thanks to its significant decrease in peak sidelobe level (PSL) \cite{15}.

In the NLFM method similar to LFM method, the signal amplitude is constant, and an optimal phase is intended to find. The stationary phase concept (SPC) is often used for NLFM method. The stationary phase concept expresses that the power spectral density (PSD) in a specific frequency is relatively large if frequency variations are small with respect to time \cite{12}. The stationary phase method has been used in \cite{16} resulted in significant reduction of PSL while the mainlobe width widens which can be neglected.

In this paper, the proposed method improves the PSL of the stationary phase method.

The remainder of the paper is organized as follows: The section 2 reviews NLFM waveform design based on the stationary phase method. NLFM waveform design based on the proposed method is explained in the section 3 in which the optimal phase is calculated initially and then the convergence of the proposed method is demonstrated using mathematical analysis. The section 4 contains simulation results. Finally, the section 5 concludes the paper.

\section{NLFM Signal Design with Stationary Phase Method}
In stationary phase method, the desired signal is defined as $x\left( t \right)$.
\begin{equation}
x\left( t \right) = a\left( t \right)\exp \left( {j\varphi \left( t \right)} \right)\;\;\;\;\;\;\; - \frac{T}{2} \le t \le \;\frac{T}{2}
\end{equation}
where $a\left( t \right)$, ${\varphi \left( t \right)}$, and $T$ are amplitude, phase, and pulse width of $x\left( t \right)$, respectively. The ${f_n}$ is instantaneous frequency of $x\left( t \right)$ at time ${t_n}$ which is determined as follow
\begin{equation}
{f_n} = \frac{1}{{2\pi }}{\varphi '\left( t_n \right)}
\end{equation}

If ${X\left( {{f_n}} \right)}$ is Fourier transform of $x\left( t \right)$ in instantaneous frequency ${f_n}$ , so based on SPC, the relation between the power spectral density and frequency variation can be denoted as following equation \cite{12}
\begin{equation}
{\left| {X\left( {{f_n}} \right)} \right|^2} \approx \;2\pi \frac{{{a^2}\left( {{t_n}} \right)}}{{\left| {\varphi ''\left( {{t_n}} \right)} \right|}}
\end{equation}

Equation (3) indicates that PSD is directly proportional to $a^2\left( t \right)$ and it is inversely proportional to value of second order derivative of ${\varphi \left( t \right)}$. In NLFM technique, signal amplitude is constant, so we consider $a\left( t \right) = A$, ($A$ = constant), and PSD only depends on the value of ${\varphi ''\left( t \right)}$. If we estimate $X(f)$ with a function such as $Z(f)$, we can rewrite (3) in frequency domain as follow \cite{12}
\begin{equation}
\theta ''\left( f \right) \approx k{\left| {{\rm}Z\left( f \right)} \right|^2},\;\;\;k = constant
\end{equation}
where $\theta\left( f \right)$ is defined as the phase of $X(f)$. If $B$ is the bandwidth of $Z(f)$, $\theta '\left( f \right)$ is obtained from the integral of $\theta ''\left( f \right)$ that can be written as follow
\begin{equation}
\theta '\left( f \right) = \mathop \int \limits_{-\frac{{B}}{2}}^f {\theta ''\left(\alpha\right)} d\alpha
\end{equation}

The group time delay function ${T_g}\left( f \right)$ is defined as following equation
\begin{equation}
{T_g}\left( f \right) =  - \frac{1}{{2\pi }}\theta '\left( f \right)
\end{equation}
If (4) and (5) are substituted in (6), so
\begin{equation}
\frac{{d{T_g}\left( f \right)}}{{df}} =  - \frac{k}{{2\pi }}{\left| {Z\left( f \right)} \right|^2}\to{T_g}\left( f \right) =  - \frac{k}{{2\pi }}\mathop \int \limits_{ - \frac{B}{2}}^f {\left| {Z\left( \alpha  \right)} \right|^2}d\alpha  + r
\end{equation}
where $r$ is constant and independent of the frequency. Also it is calculated by using the following boundary conditions
\begin{equation}
{T_g}\left( {B/2} \right) = T/2,\;\;{T_g}\left( { - B/2} \right) =  - T/2
\end{equation}

Now, the frequency function of time can be determined as following equation \cite{17}
\begin{equation}
f\left( t \right) = T_g^{ - 1}\left( f \right)
\end{equation}

Finding the inverse group time delay function is not always easy; and in some cases it should be carried
out numerically. Eventually, the phase of $x\left( t \right)$ can be calculated by integral of frequency function as follow
\begin{equation}
 \varphi \left( t \right) = 2\pi\int\limits_{-\frac{{T}}{2}}^t f\left( \alpha  \right)d\alpha
\end{equation}

This method was applied in \cite{16}. In section 4, the results are compared against the proposed method.

\section{NLFM Signal Design with Proposed Method}

Our proposed method is based on a constrained optimization problem. First, a window as the initial window is considered, then by solving the constrained optimization problem, the desired signal is aimed to be found. This method is performed iteratively and the phase obtained from the stationary phase method \cite{16} is used to start. To guarantee the convergence of the proposed method, triangle inequality and mathematical analysis are used to demonstrate the minimum error decrement in each iteration. Additionally, the minimum error value is positive expressing convergence due in large to the fact that a positive nonincreasing sequence certainly converges.

\subsection{Optimal Phase}
In the proposed method, to obtain the phase of the desired signal, first, an initial window such as ${W_{initial}}$ is considered where $\left| {Y\left( f \right)} \right|$ is its root. Difference between $\left| {Y\left( f \right)} \right|$ and the amplitude of the Fourier transform of the desired signal is defined as the error. Since in NLFM signals, amplitude is constant, so our goal is to minimize the error with the constraint of being unit the amplitude of $x\left( t \right)$ for $| t | \le T/2$, therefore we try to minimize the following equation
\begin{equation}
\begin{array}{cl}
\min\limits_{X(f)} 
&E = \displaystyle\int \limits_{ - \frac{B}{2}}^{\frac{B}{2}} {\Big| {\left| {Y\left( f \right)} \right| - \left|X\left( f \right)\right|} \Big|^2}df\\
\text{s.t.} &
\begin{cases}
| {x(t)} |^2 = 1 &| t | \le T/2\\
x(t) = 0&| t | > T/2
\end{cases}
\end{array}
\end{equation}
If two complex numbers come close to each other, then it can be concluded that the values of their amplitudes also close together, so if the following equation is reduced, then the error can be reduced, which is expressed in the phase matching problems \cite{18,19,20}
\begin{equation}
\begin{array}{cl}
\min\limits_{\theta (f),X(f)} 
&E = \displaystyle\int \limits_{ - \frac{B}{2}}^{\frac{B}{2}} {\Big| {\left| {Y\left( f \right)} \right|\exp (j\theta (f)) - X\left( f \right)} \Big|^2}df\\
\text{s.t.} &
\begin{cases}
| {x(t)} |^2 = 1 &| t | \le T/2\\
x(t) = 0&| t | > T/2
\end{cases}
\end{array}
\end{equation}
Assume ${Y_\theta }\left( f \right) = \left| {Y\left( f \right)} \right|\exp \left( {j\theta \left( f \right)} \right)$ and $f = kB/(K - 1)$, then the error equation can be written in the discrete form as follow
\begin{equation}
E =  \sum \limits_{k = 0}^{K - 1} {\left| {{Y_\theta }\left( k \right) - X\left( k \right)} \right|^2} 
\end{equation}
where ${X\left( k \right) = \mathop \sum \limits_{n = 0}^{N - 1} x\left( n \right)\exp \left( { - j\frac{{2\pi kn}}{K}} \right)}$.
Consider equations in vector space and assume ${\bf{x}} = {\left[ {x\left( 0 \right),\ x\left( 1 \right), \dots,\ x\left( {N - 1} \right)} \right]^T}$ and ${{\left[ {\bf{W}} \right]_{k,n}} = \exp \left( { - j\frac{{2\pi kn}}{K}} \right)}$, so
\begin{equation}
\left[ {\begin{array}{*{20}{c}}
{\begin{array}{*{20}{c}}
{X\left( 0 \right)}\\
\end{array}}\\
{\begin{array}{*{20}{c}}
 \vdots \\
{X\left( {K - 1} \right)}
\end{array}}
\end{array}} \right] = {\bf{Wx}}
\end{equation}
If we assume ${{\bf{Y}}_{\bf{\theta }}} = {\left[ {{Y_\theta }\left( 0 \right),\ {Y_\theta }\left( 1 \right), \ldots.\ {Y_\theta }\left( {K - 1} \right)} \right]^T}$, the (12) is rewritten as follow
\begin{equation}
\begin{array}{rl}
\min\limits_{\bf \uptheta, \bf{x}} & {\left( {{{\bf{Y}}_{\bf{\theta }}} - {\bf{Wx}}} \right)^H}\left( {{{\bf{Y}}_{\bf{\theta }}} - {\bf{Wx}}} \right)\\
\text{s.t.}&{\left| {x\left( n \right)} \right|^2} = 1,\quad n=1,\dots,N-1
\end{array}
\end{equation}
Using Lagrangian method, we solve the obtained constrained optimization problem \cite{21}.
\begin{align}
J &= {\left( {{{\bf{Y}}_{\bf{\theta }}} - {\bf{Wx}}} \right)^H}\left( {{{\bf{Y}}_{\bf{\theta }}} - {\bf{Wx}}} \right) + \displaystyle\sum_{n=0}^{N-1}{\lambda _n}\;{\left| {x\left( n \right)} \right|^2} \nonumber\\
&= {{\bf{Y}}_{\bf{\theta }}}^H{{\bf{Y}}_{\bf{\theta }}} {-} {{\bf{Y}}_{\bf{\theta }}}^H{\bf{Wx}} {-} {{\bf{x}}^H}{{\bf{W}}^H}{{\bf{Y}}_{\bf{\theta }}} {+} {{\bf{x}}^H}{{\bf{W}}^H}{\bf{Wx}} {+} {{\bf{x}}^H}{\bf{\Lambda }}{\bf{x}}
\end{align}
where ${\lambda _n}$ is Lagrange multiplier and ${\bf{\Lambda }} = \text{diag}({\lambda _0},\;{\lambda _1},\;...,{\lambda _{N - 1}})$. The symbol diag(.) is the diagonal matrix which the entries outside the main diagonal are all zero. Because of the orthogonality of the columns of the matrix ${\bf{W}}$, the value of ${{\bf{W}}^H}{\bf{W}}$ is equal to $K{{\bf{I}}_N}$ where ${{\bf{I}}_N}$ is identity matrix of size $N$. We take $J$ derivative with respect to ${\bf{x}}$.
\begin{gather}
\cfrac{{\partial J}}{{\partial {\bf{x}}}} =  - {({{\bf{W}}^H}{{\bf{Y}}_{\bf{\theta }}})^*} + K{{\bf{I}}_N}{{\bf{x}}^*} + {\bf{\Lambda }}{{\bf{x}}^*} = 0\nonumber\\
 \Rightarrow {\bf{x}} = {(K{{\bf{I}}_N} + {{\bf{\Lambda }}^*})^{ - 1}}{{\bf{W}}^H}{{\bf{Y}}_{\bf{\theta }}}
\end{gather}
where * denotes the complex conjugate, ${(K{{\bf{I}}_N} + {{\bf{\Lambda }}^*})^{ - 1}}$ is the inverse matrix of $K{{\bf{I}}_N} + {{\bf{\Lambda }}^*}$ and calculated as follow
\begin{equation}
{(K{{\bf{I}}_N} + {{\bf{\Lambda }}^*})^{ - 1}} = 
\text{diag}\left({\frac{1}{{K + \lambda _0^*}}},\ \dots,\ {\frac{1}{{K + \lambda _{N-1}^*}}}\right)
\end{equation}
Since the constraints of the optimization problem are real, then the Lagrange multipliers ${\lambda _n}$ are real and the vector ${\bf{x}}$ is expressed as follow
\begin{equation}
\begin{bmatrix}
x(0)\\
 \vdots \\
x(N - 1)
\end{bmatrix}=
\begin{bmatrix}
\frac{1}{{K + {\lambda _0}}}\displaystyle\sum\limits_{k = 0}^{K - 1} {{{\left[ {{{\bf{W}}^H}} \right]}_{1,k + 1}}{Y_\theta }\left( k \right)} \\
 \vdots \\
\frac{1}{{K + {\lambda _{N - 1}}}}\displaystyle\sum\limits_{k = 0}^{K - 1} {{{\left[ {{{\bf{W}}^H}} \right]}_{N,k + 1}}{Y_\theta }\left( k \right)} 
\end{bmatrix}
\end{equation}
Due to the constraints of the problem, the values ${\lambda _n}$ must be such that the square of the amplitude of each coefficient of ${\bf{x}}$ is equal to one, so
\begin{gather}
{\left| {\frac{1}{{K + {\lambda _n}}}\sum\limits_{k = 0}^{K - 1} {{{\left[ {{{\bf{W}}^H}} \right]}_{n+1,k + 1}}{Y_\theta }\left( k \right)} } \right|^2} = 1\nonumber\\
 \Rightarrow {\lambda _n} =  \pm \left| {\sum\limits_{k = 0}^{K - 1} {{{\left[ {{{\bf{W}}^H}} \right]}_{n+1,k + 1}}{Y_\theta }\left( k \right)} } \right| - K
\end{gather}
Therefore, the vector ${\bf{x}}$ is calculated as follow.
\begin{equation}
{\bf{x}} =
\left[ \begin{array}{c}
\frac{{\sum\limits_{k = 0}^{K - 1} {{{\left[ {{{\bf{W}}^H}} \right]}_{1,k + 1}}{Y_\theta }\left( k \right)} }}{{\left| {\sum\limits_{k = 0}^{K - 1} {{{\left[ {{{\bf{W}}^H}} \right]}_{1,k + 1}}{Y_\theta }\left( k \right)} } \right|}}\\
 \vdots \\
\frac{{\sum\limits_{k = 0}^{K - 1} {{{\left[ {{{\bf{W}}^H}} \right]}_{N,k + 1}}{Y_\theta }\left( k \right)} }}{{\left| {\sum\limits_{k = 0}^{K - 1} {{{\left[ {{{\bf{W}}^H}} \right]}_{N,k + 1}}{Y_\theta }\left( k \right)} } \right|}}
\end{array} \right] = {{\bf{\Lambda }}_1}{{\bf{W}}^H}{{\bf{Y}}_{\bf{\theta }}}
\end{equation}
where ${{\bf{\Lambda }}_1}$ is a $N\times N$ matrix as follow
\begin{equation}
\left[{{\bf{\Lambda }}_1}\right]_{i,j} = 
\begin{cases}
{{{\left| {\sum\limits_{k = 0}^{K - 1} {{{\left[ {{{\bf{W}}^H}} \right]}_{i,k + 1}}{Y_\theta }\left( k \right)} } \right|^{-1}}}}, &
i=j\\
0,&i\neq j
\end{cases}
\end{equation}
To achieve the desired signal, the proposed method is performed as an iterative algorithm; therefore, in respect to (21), in r-th iteration, the desired signal will be as follow
\begin{equation}
{{\bf{x}}^{(r)}} = {\bf{\Lambda }}_1^{(r - 1)}{{\bf{W}}^H}{\bf{Y}}_{\bf{\theta }}^{(r - 1)}
\end{equation}
${{\bf{\uptheta }}^{\left( r \right)}}$ is the phase of $X\left( f \right)$ in r-th iteration, which can be calculated as follow
\begin{equation}
{{\bf{\uptheta }}^{\left( r \right)}} = {\rm{phase}}({\bf{W}}{{\bf{x}}^{(r)}})
\end{equation}
By calculating the ${{\bf{\uptheta }}^{\left( r \right)}}$ value, vector ${\bf{Y}}_{\bf{\theta }}^{(r)}$ and then the matrix ${\bf{\Lambda }}_1^{(r)}$ is calculated.
\begin{equation}
{\bf{Y}}_{\bf{\theta }}^{(r)} = \left[ {\begin{array}{*{20}{c}}
{\left| {Y\left( 0 \right)} \right|\exp (j{\theta ^{(r)}}(0))}\\
 \vdots \\
{\left| {Y\left( {K - 1} \right)} \right|\exp (j{\theta ^{(r)}}(K - 1))}
\end{array}} \right]
\end{equation}
\begin{equation}
\left[{{\bf{\Lambda }}_1^{(r)}}\right]_{i,j} = 
\begin{cases}
{{{\left| {\sum\limits_{k = 0}^{K - 1} {{{\left[ {{{\bf{W}}^H}} \right]}_{i,k + 1}}{Y_\theta^{(r)} }\left( k \right)} } \right|^{-1}}}}, &
i=j\\
0,&i\neq j
\end{cases}
\end{equation}
To start the algorithm, we set the ${{\bf{\uptheta }}^{\left( 0 \right)}}$ equal to the phase value obtained from the stationary phase method for the Fourier transform of the NLFM signal \cite{16}. Thus, by repeating the algorithm, we obtain the desired NLFM signal, and because the amplitude of NLFM signal is constant, so we can multiply the amplitude of the obtained signal in the constant coefficient $A$.

\subsection{Convergence of the Proposed Method}
With respect to the obtained value for the vector ${\bf{x}}$ in (21) substituted in (15), the minimum value of $E$ is calculated as follow.
\begin{equation}
\begin{array}{l}
{E_{\text{Min}}} 
 = {{\bf{Y}}_{\bf{\theta }}}^H\left({{\bf{I}}_K} - {\bf{W}}(2{{\bf{\Lambda }}_1} - K{\bf{\Lambda }}_1^2){{\bf{W}}^H}\right){{\bf{Y}}_{\bf{\theta }}}
\end{array}
\end{equation}
In (27), ${\bf{P}} = {\bf{W}}(2{{\bf{\Lambda }}_1} - K{\bf{\Lambda }}_1^2){{\bf{W}}^H}$ and ${\bf{A}} = {{\bf{I}}_K} - {\bf{W}}(2{{\bf{\Lambda }}_1} - K{\bf{\Lambda }}_1^2){{\bf{W}}^H}$ are the projection and orthogonal complement matrices, respectively \cite{22}. The minimum error in r-th iteration is as follow
\begin{equation}
\Scale[0.845]{
E_{\text{Min}}^{(r)} = {\left({{\bf{Y}}_{\bf{\theta }}}^{(r-1)}\right)^H}
\left({{\bf{I}}_K} - {\bf{W}}\left(2{\bf{\Lambda }}_1^{(r-1)} - K{({\bf{\Lambda }}_1^{(r-1)})^2}\right){{\bf{W}}^H}\right)
{{\bf{Y}}_{\bf{\theta }}}^{(r-1)}}
\end{equation}
For convergence of the proposed method, the error value must be reduced with increasing of iterations. In other words, for the convergence of the proposed method, the following inequality should be satisfied.
\begin{equation}
0 \le E_{\text{Min}}^{(r + 1)} \le E_{\text{Min}}^{(r)}
\end{equation}
To prove (29), we use (12) to represent the minimum error in r-th iteration as
\begin{equation}
\begin{array}{l}
E_{\text{Min}}^{^{(r)}} =  \displaystyle\int \limits_{ - \frac{B}{2}}^{\frac{B}{2}}\Big| {\left| {Y\left( f \right)} \right|\exp (j{\theta ^{(r - 1)}}(f)) - {X^{(r)}}\left( f \right)} \Big|^2df
\end{array}
\end{equation}
According to the triangle inequality, we can write
\begin{equation}
\begin{array}{ll}
E_{\text{Min}}^{^{(r)}} 
& \ge \displaystyle\int \limits_{-\frac{B}{2}}^{\frac{B}{2}} \left||Y( f )| - | X^{(r)}(f)| \right|^2df\\
&=\displaystyle\int \limits_{-\frac{B}{2}}^{\frac{B}{2}}
 \Big| {\exp (j{\theta ^{(r)}}(f))} \Big|^2
\left||Y( f )| - | X^{(r)}(f)| \right|^2df\\
&= \displaystyle\int \limits_{ - \frac{B}{2}}^{\frac{B}{2}}\Big| {\left| {Y\left( f \right)} \right|\exp (j{\theta ^{(r)}}(f)) - {X^{(r)}}\left( f \right)} \Big|^2df
\end{array}
\end{equation}
On the other hand, since $E_{\text{Min}}^{(r + 1)}$ is the minimum value of error in $(r {+} 1)$-th iteration, the following equation is satisfied.
\begin{equation}
\begin{array}{ll}
E_{\text{Min}}^{^{(r + 1)}} &= 
\displaystyle\int \limits_{ - \frac{B}{2}}^{\frac{B}{2}}\Big| {\left| {Y\left( f \right)} \right|\exp (j{\theta ^{(r)}}(f)) - {X^{(r+1)}}\left( f \right)} \Big|^2df\\
&\le\displaystyle\int \limits_{ - \frac{B}{2}}^{\frac{B}{2}}\Big| {\left| {Y\left( f \right)} \right|\exp (j{\theta ^{(r)}}(f)) - {X^{(r)}}\left( f \right)} \Big|^2df
\end{array}
\end{equation}
From the comparison of the two equations (31) and (32), we conclude $E_{\text{Min}}^{^{(r)}} \ge E_{\text{Min}}^{^{(r + 1)}}$. Since we considered two arbitrary successive iterations, so (29) is satisfied for each two arbitrary successive iterations.
Since $E_{\text{Min}}^{(r)}$ is a positive nonincreasing sequence, convergence of the proposed method is guaranteed.

\section{Simulation Results}
The proposed method is performed for six initial windows of Raised-Cosine, Taylor, Chebyshev, Gaussian, Poisson, and Kaiser. Table 1 shows the formula of the selected windows, the group time delay functions, and their constant parameters. As already mentioned, the group time delay function for some windows is numerically calculated. In the Table 1, erf(.) and sgn(.) are defined as the error and sign functions, respectively. The design parameters such as $\text{bandwidth ($B$)}$, pulse width ($T$), and sampling rate are considered equal to $\text{100 MHz}$, 2.5 $\mu$s, and 1 GHz, respectively. Fig. 1 and $\text{Fig. 2}$ show the autocorrelation functions of the designed signals using the stationary phase method and the proposed method for the six selected windows, respectively. Also, Fig. 3 shows the minimum error for the six selected windows in different iterations.

\begin{figure*}[!h]
	\centering
	\mbox{\subfloat[]{\label{subfig10:a} 	\def\big{\includegraphics[width=0.48\textwidth,height=6.25cm]{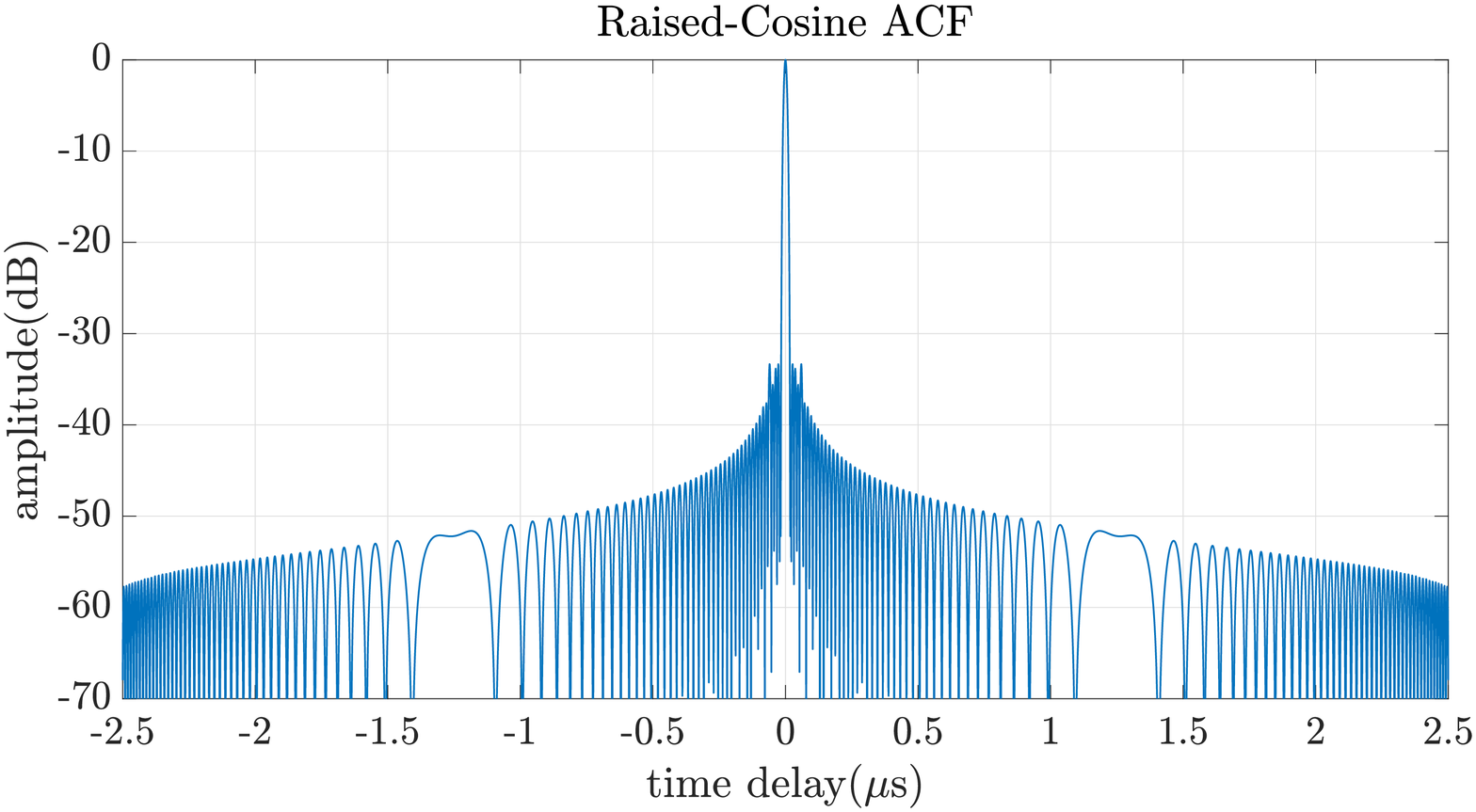}}
			\def\little{\includegraphics[height=2.4cm]{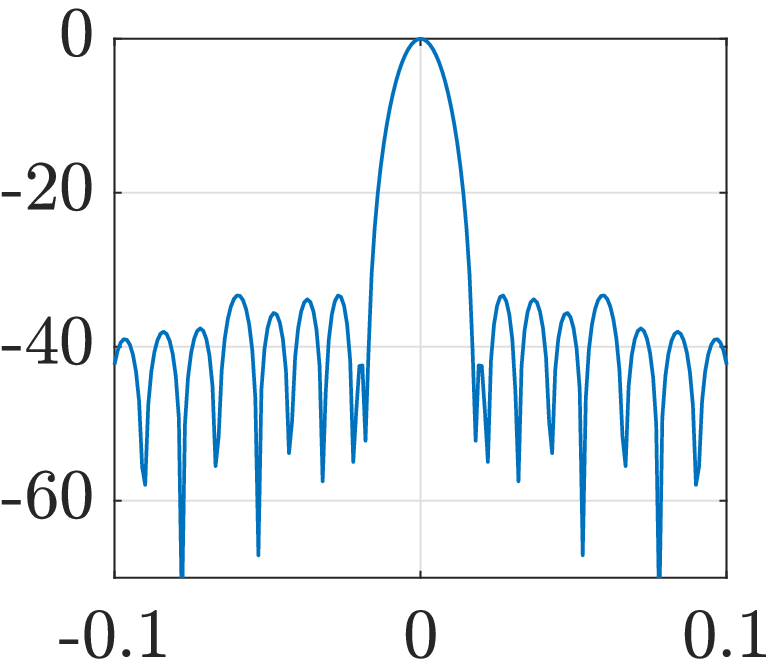}}
			\def\stackalignment{r}
			\topinset{\little}{\big}{15pt}{24pt}}}\vspace{24pt}
	\mbox{\subfloat[]{\label{subfig10:b} 	\def\big{\includegraphics[width=0.48\textwidth,height=6.25cm]{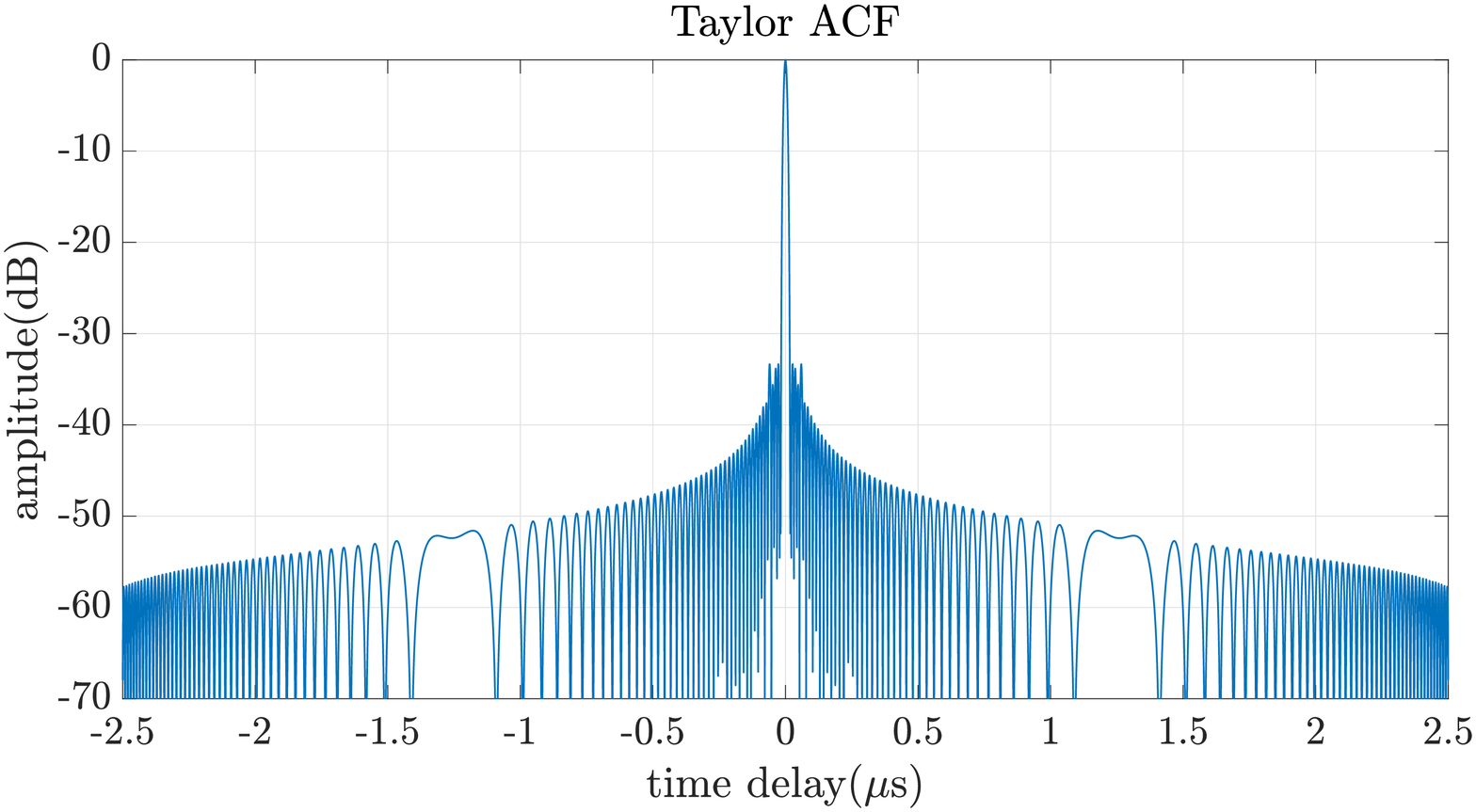}}
			\def\little{\includegraphics[height=2.4cm]{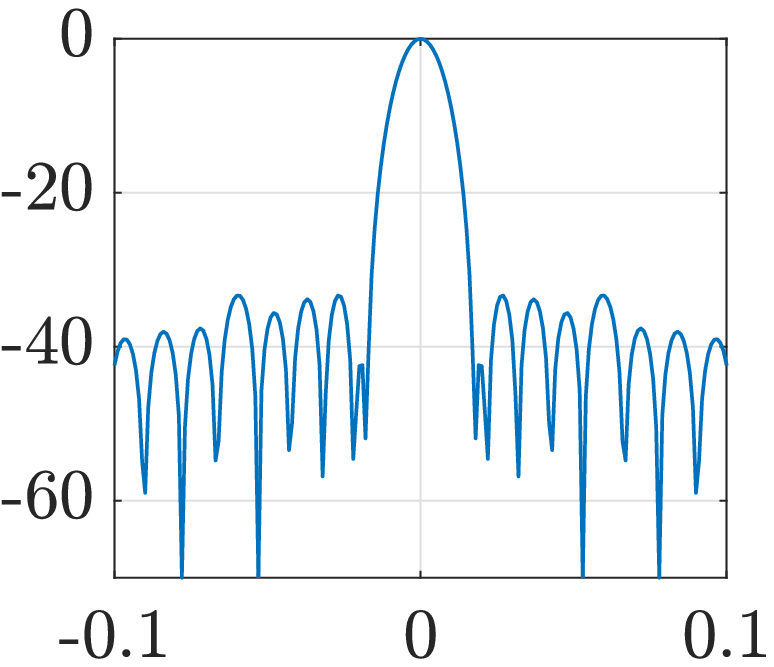}}
			\def\stackalignment{r}
			\topinset{\little}{\big}{15pt}{24pt}}}
	\mbox{\subfloat[]{\label{subfig10:c} 	\def\big{\includegraphics[width=0.48\textwidth,height=6.25cm]{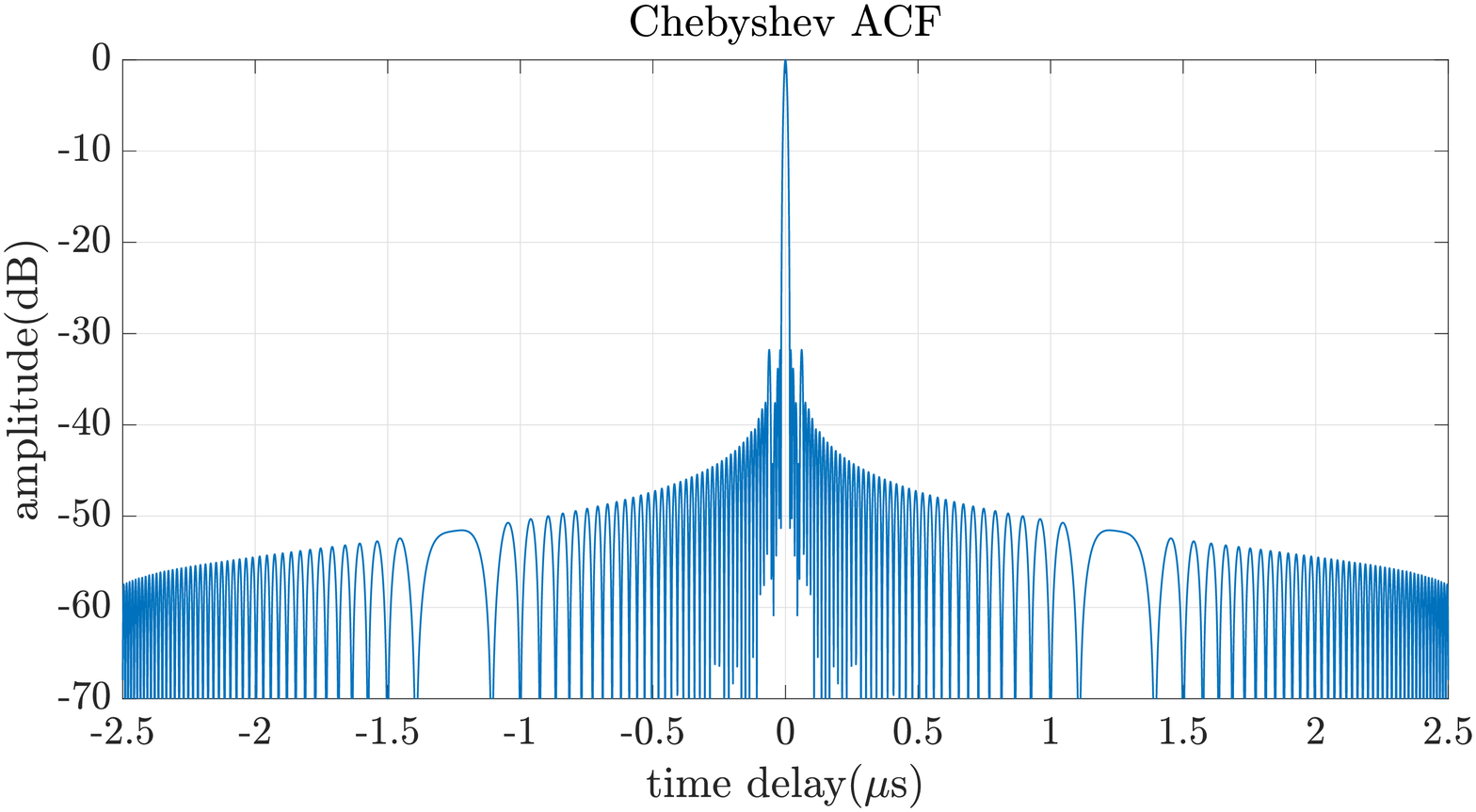}}
			\def\little{\includegraphics[height=2.4cm]{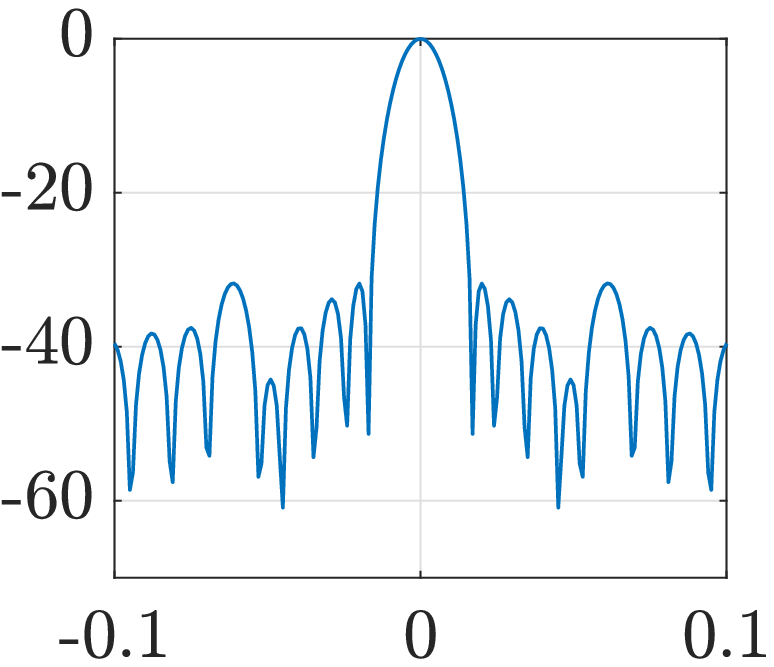}}
			\def\stackalignment{r}
			\topinset{\little}{\big}{15pt}{24pt}}}\vspace{24pt}
	\mbox{\subfloat[]{\label{subfig10:d} 	\def\big{\includegraphics[width=0.48\textwidth,height=6.25cm]{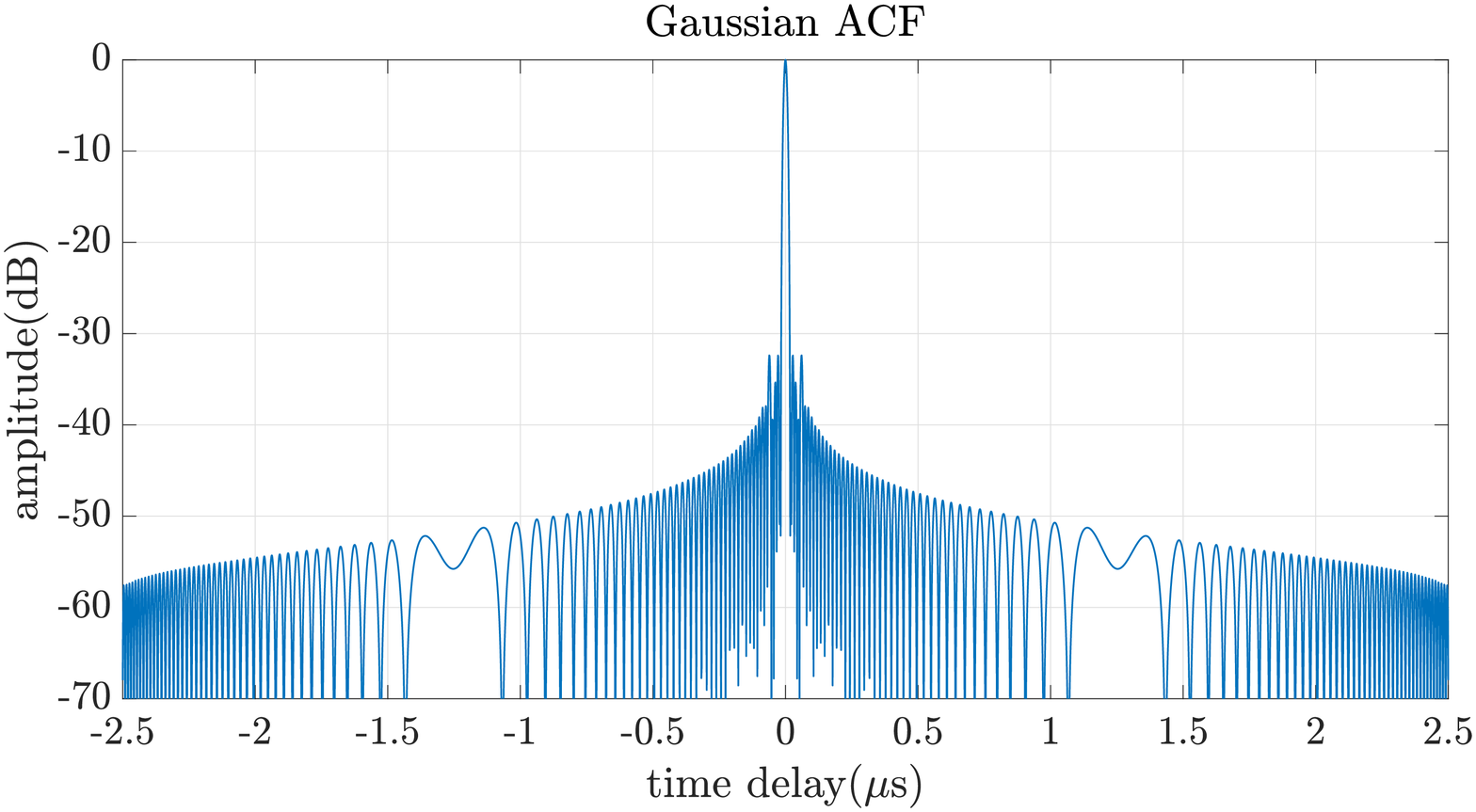}}
			\def\little{\includegraphics[height=2.4cm]{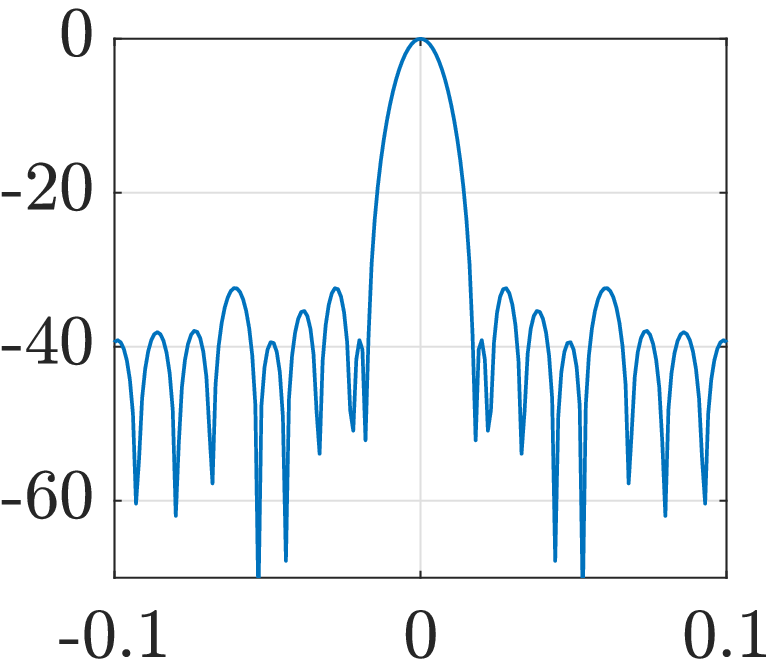}}
			\def\stackalignment{r}
			\topinset{\little}{\big}{15pt}{24pt}}}
          \mbox{\subfloat[]{\label{subfig10:d} 	\def\big{\includegraphics[width=0.48\textwidth,height=6.25cm]{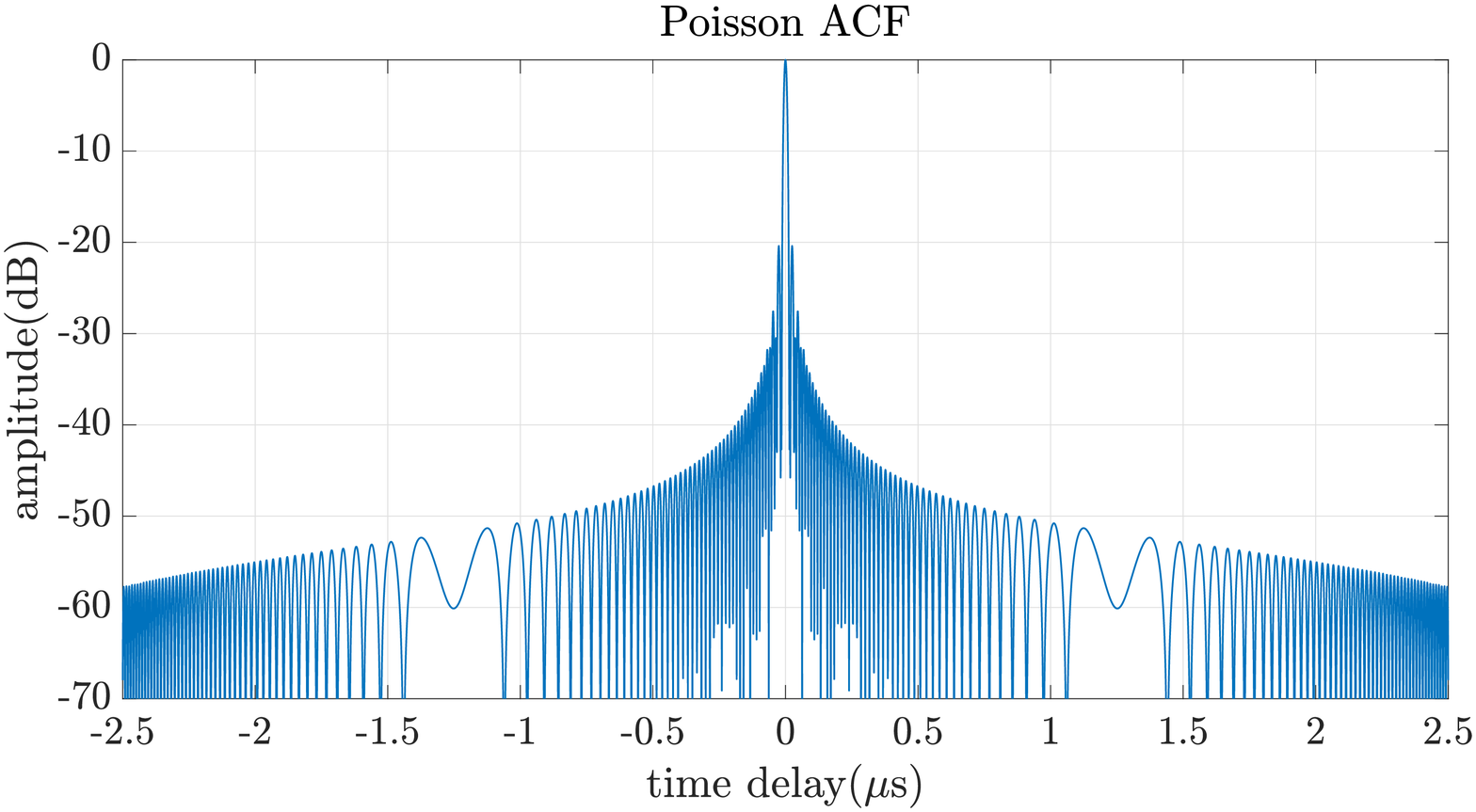}}
			\def\little{\includegraphics[height=2.4cm]{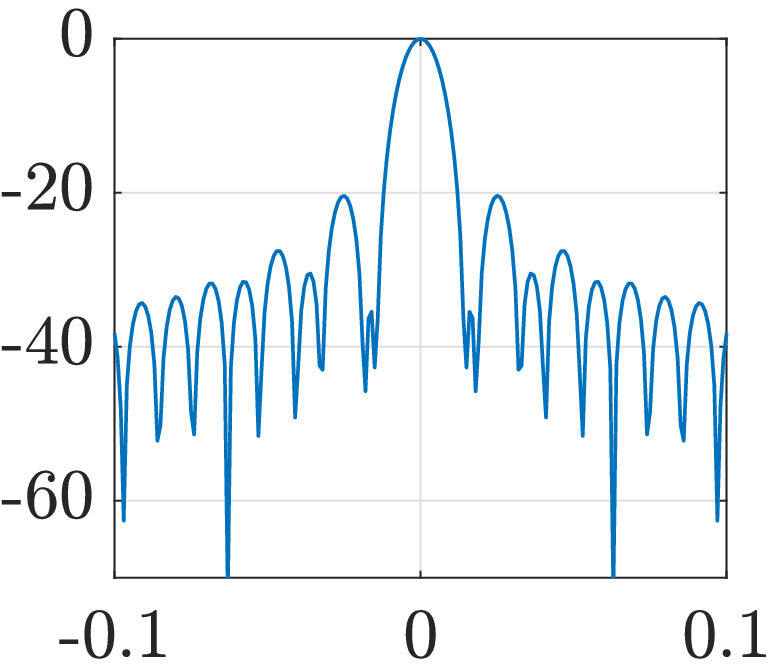}}
			\def\stackalignment{r}
			\topinset{\little}{\big}{15pt}{24pt}}}
          \mbox{\subfloat[]{\label{subfig10:d} 	\def\big{\includegraphics[width=0.48\textwidth,height=6.25cm]{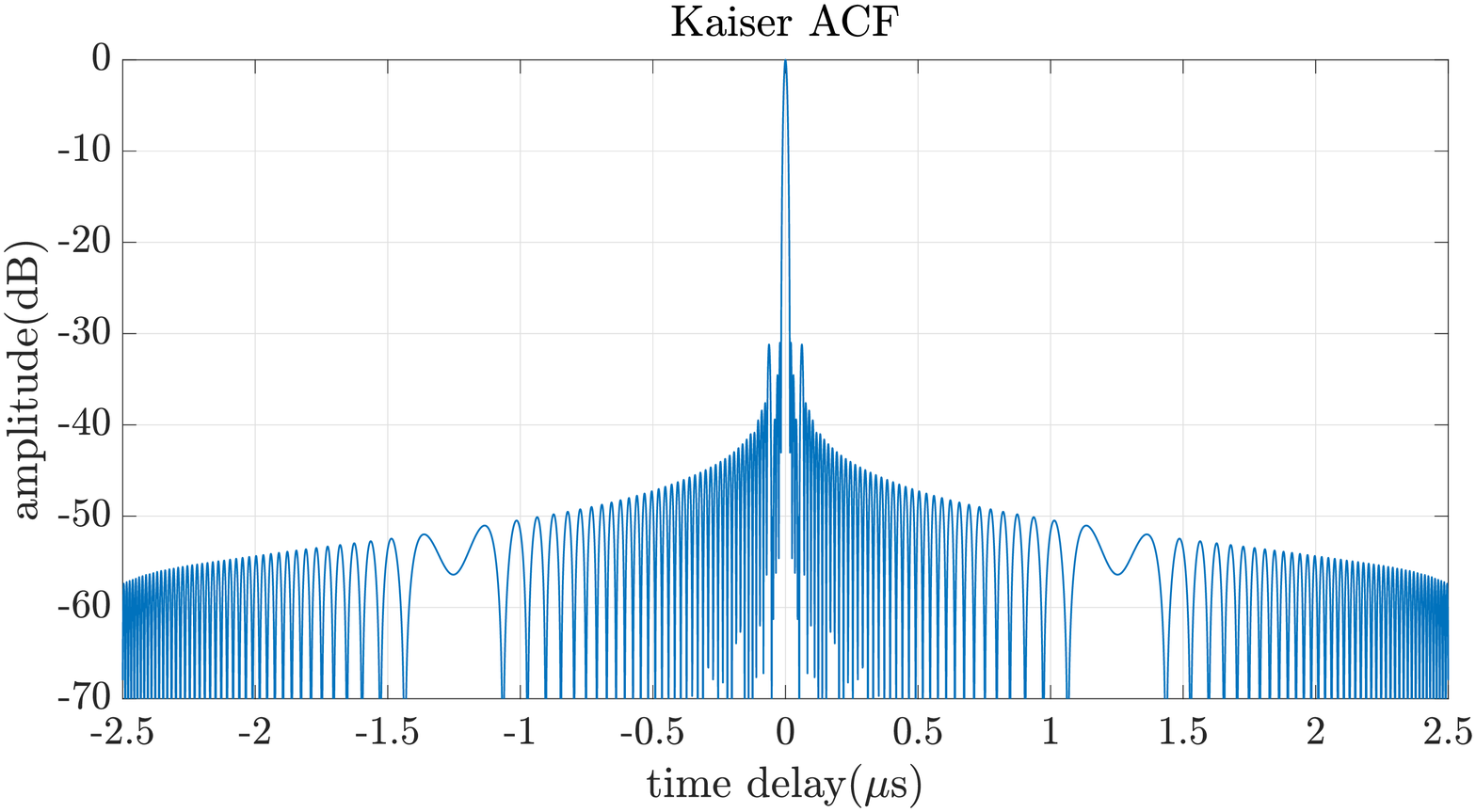}}
			\def\little{\includegraphics[height=2.4cm]{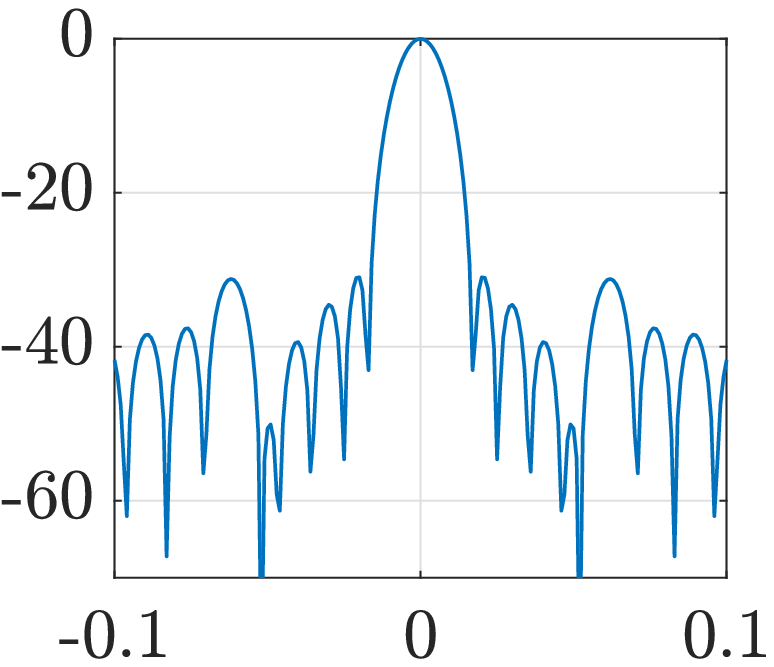}}
			\def\stackalignment{r}
			\topinset{\little}{\big}{15pt}{24pt}}}
	\caption{The autocorrelation functions of the designed
signals using the stationary phase method for the six windows of Raised-Cosine, Taylor, Chebyshev, Gaussian, Poisson, and Kaiser.}
	\label{fig:fig10}
\end{figure*}
\begin{figure*}[!h]
	\centering
	\mbox{\subfloat[]{\label{subfig10:a} 	\def\big{\includegraphics[width=0.48\textwidth,height=6.25cm]{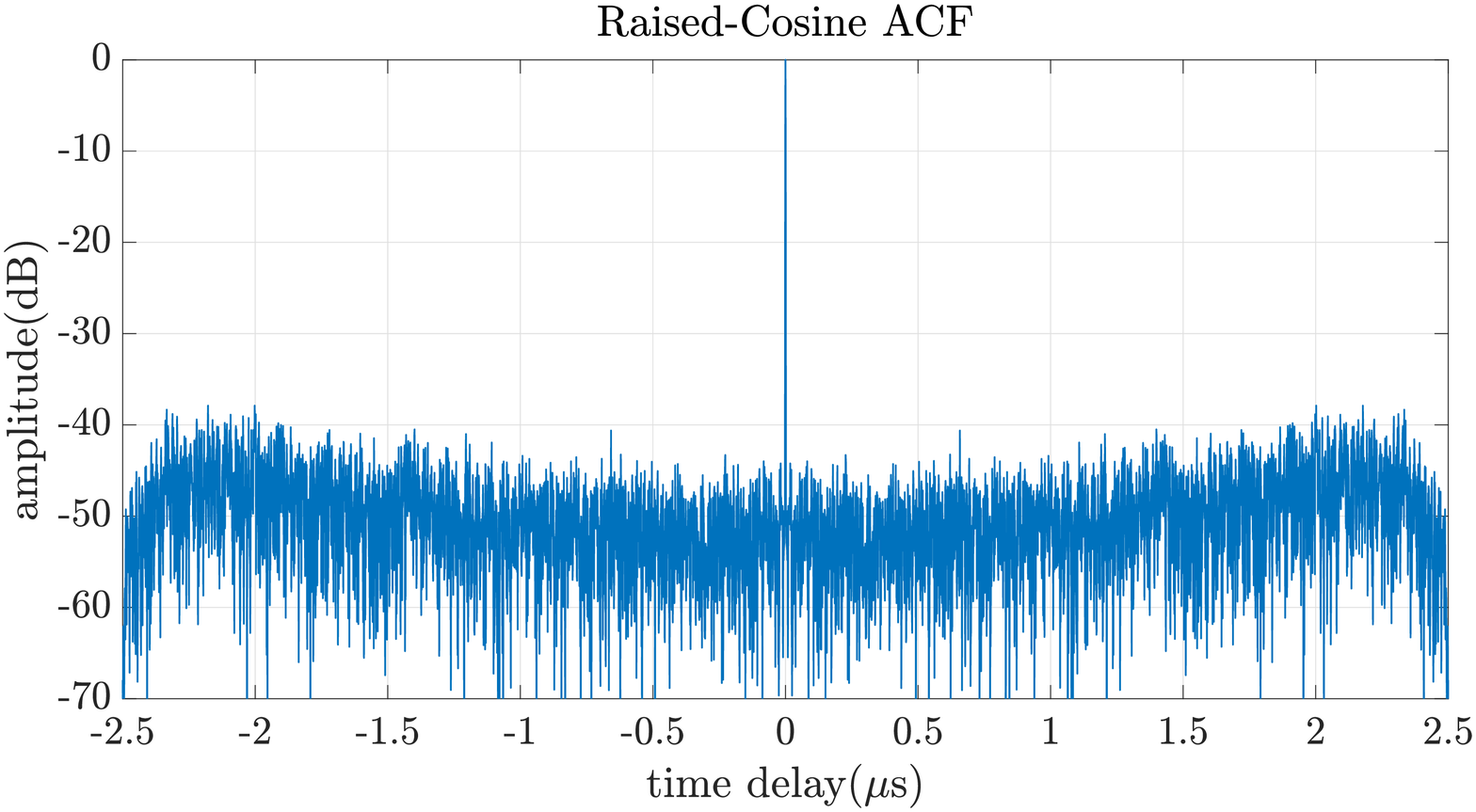}}
			\def\little{\includegraphics[height=2.4cm]{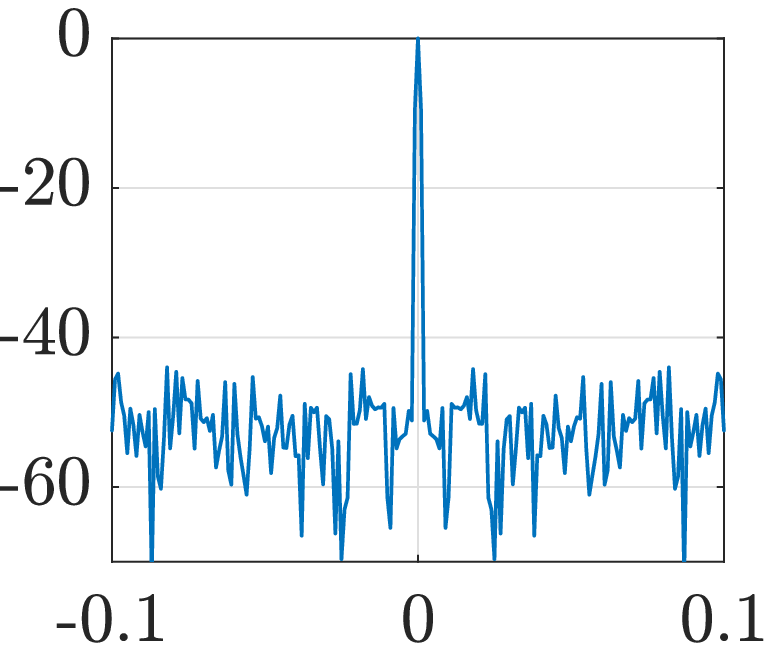}}
			\def\stackalignment{r}
			\topinset{\little}{\big}{15pt}{24pt}}}\vspace{24pt}
	\mbox{\subfloat[]{\label{subfig10:b} 	\def\big{\includegraphics[width=0.48\textwidth,height=6.25cm]{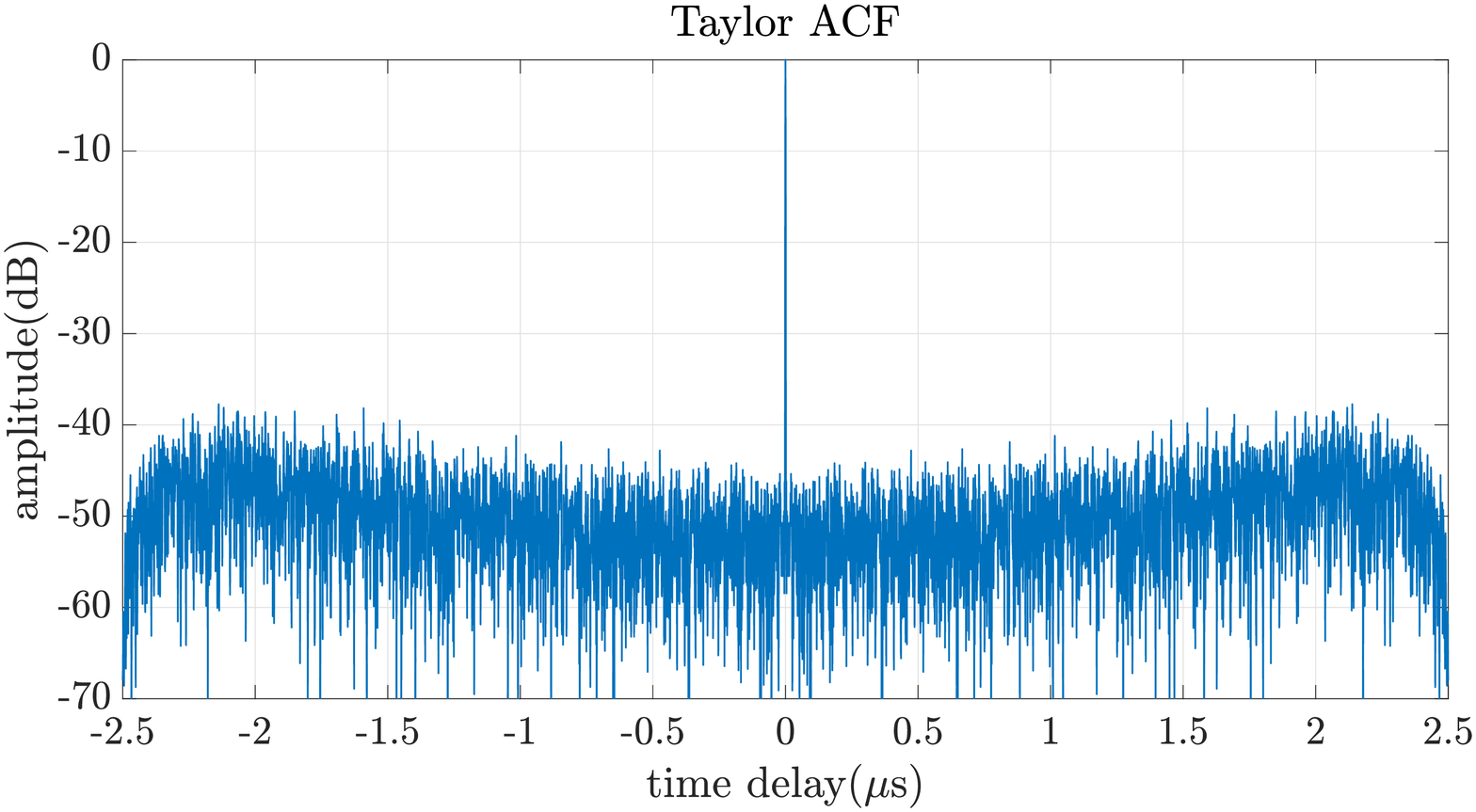}}
			\def\little{\includegraphics[height=2.4cm]{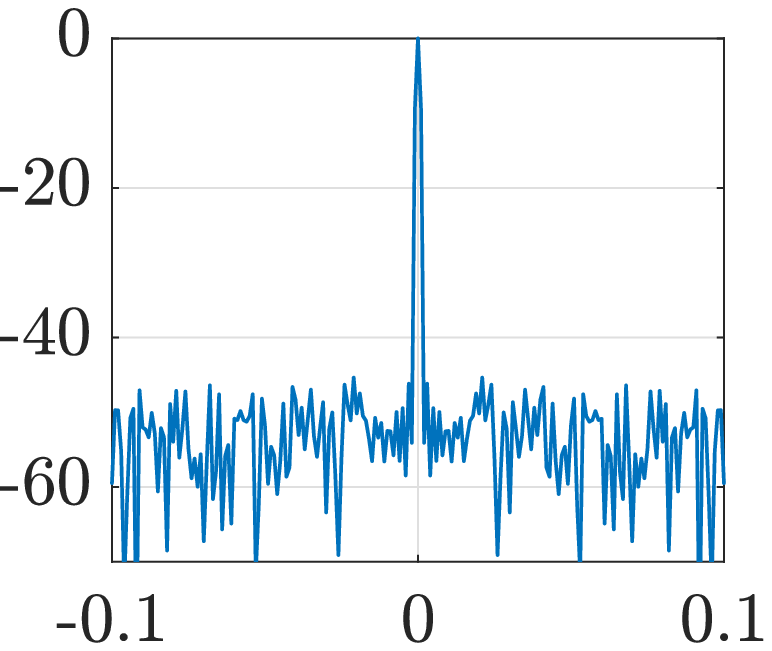}}
			\def\stackalignment{r}
			\topinset{\little}{\big}{15pt}{24pt}}}
	\mbox{\subfloat[]{\label{subfig10:c} 	\def\big{\includegraphics[width=0.48\textwidth,height=6.25cm]{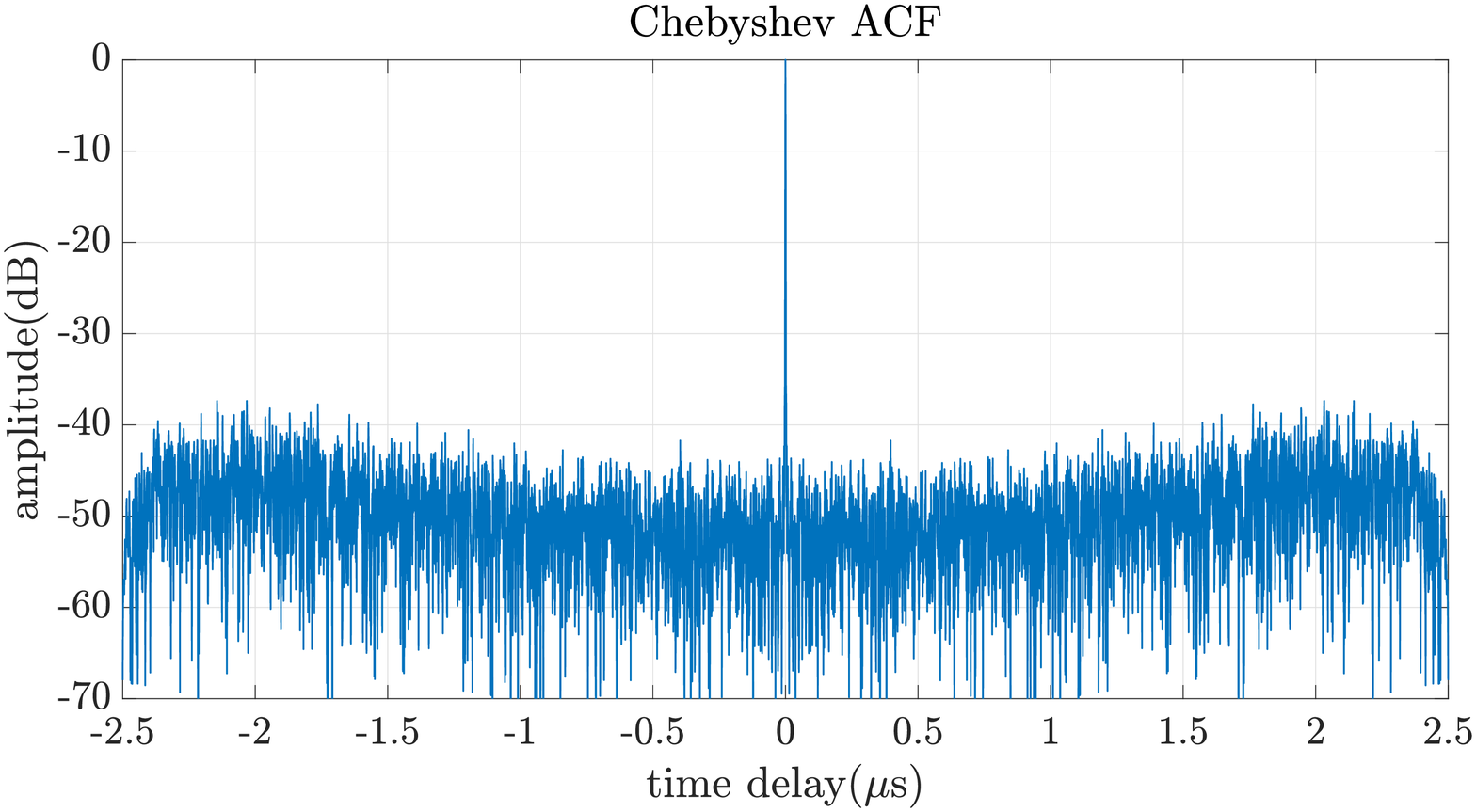}}
			\def\little{\includegraphics[height=2.4cm]{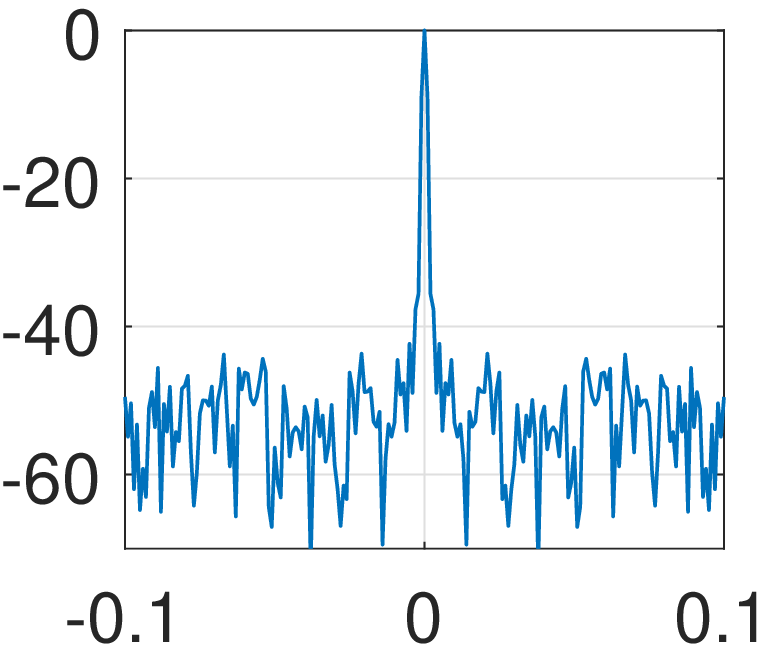}}
			\def\stackalignment{r}
			\topinset{\little}{\big}{15pt}{24pt}}}\vspace{24pt}
	\mbox{\subfloat[]{\label{subfig10:d} 	\def\big{\includegraphics[width=0.48\textwidth,height=6.25cm]{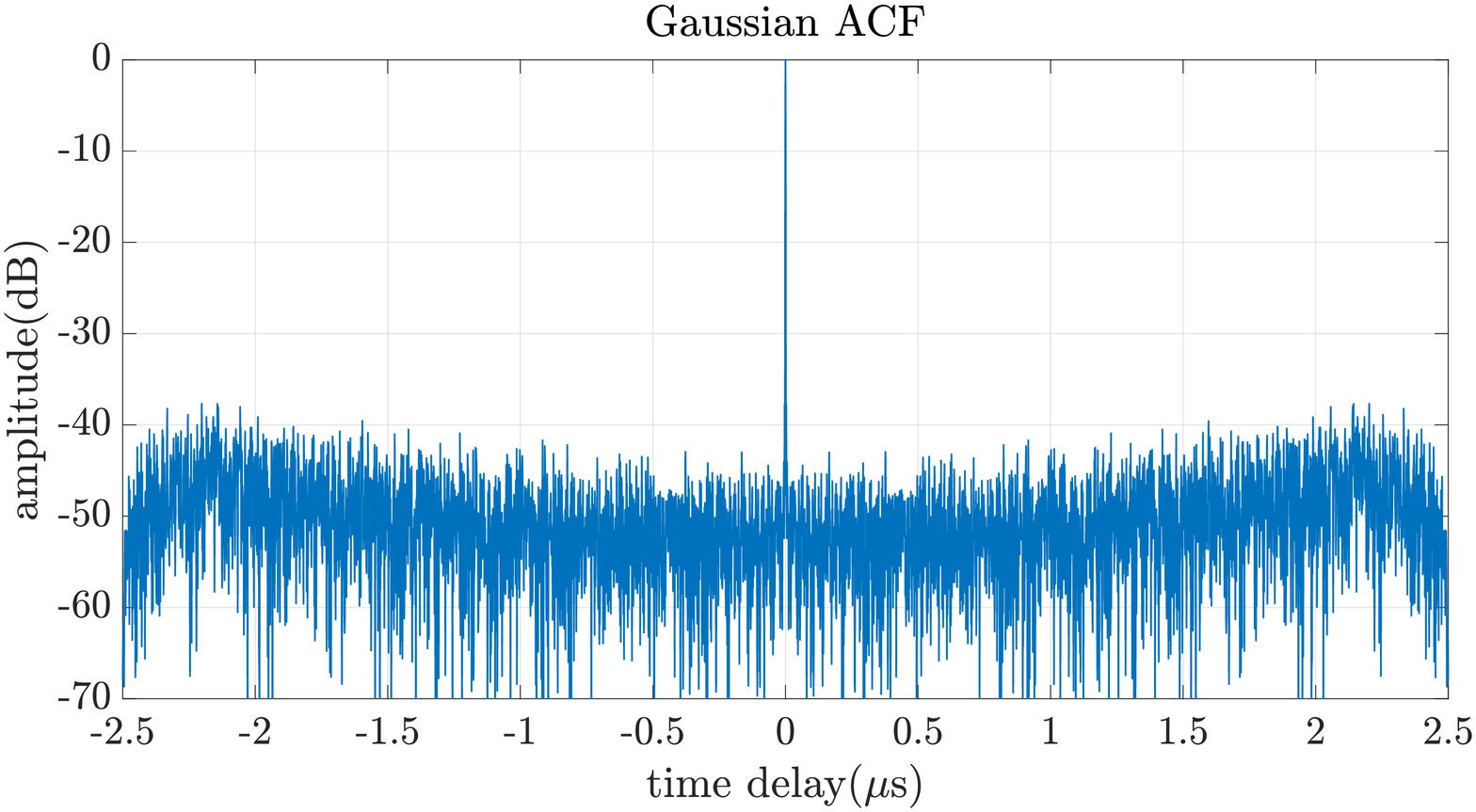}}
			\def\little{\includegraphics[height=2.4cm]{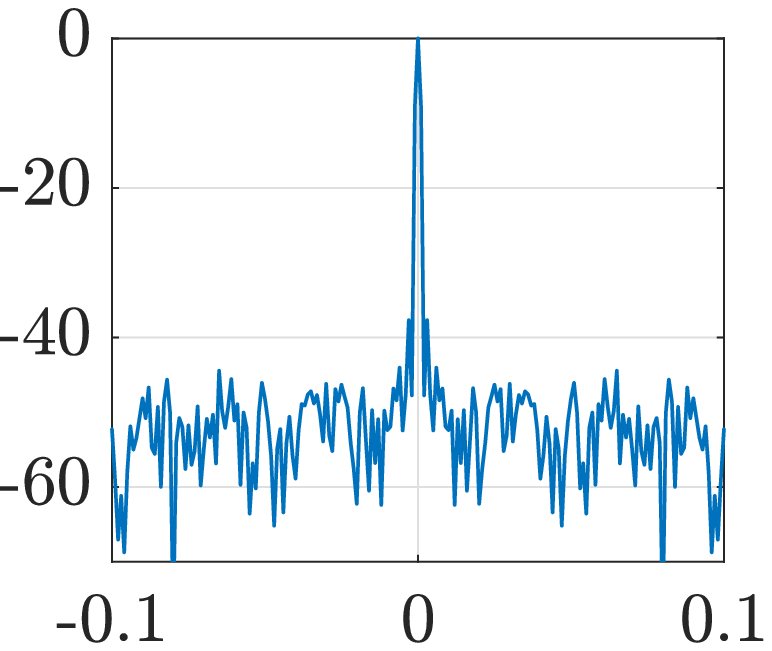}}
			\def\stackalignment{r}
			\topinset{\little}{\big}{15pt}{24pt}}}
          \mbox{\subfloat[]{\label{subfig10:d} 	\def\big{\includegraphics[width=0.48\textwidth,height=6.25cm]{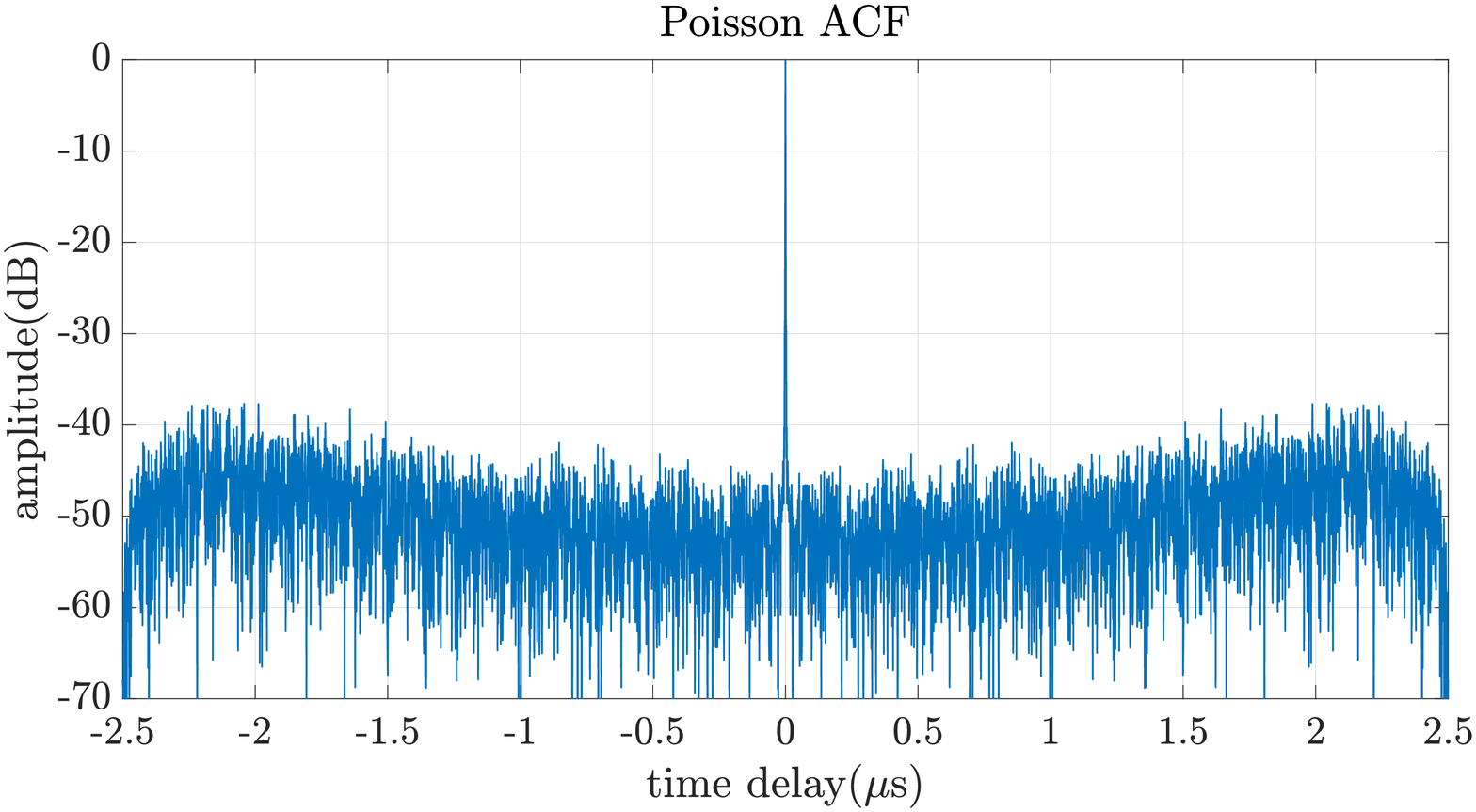}}
			\def\little{\includegraphics[height=2.4cm]{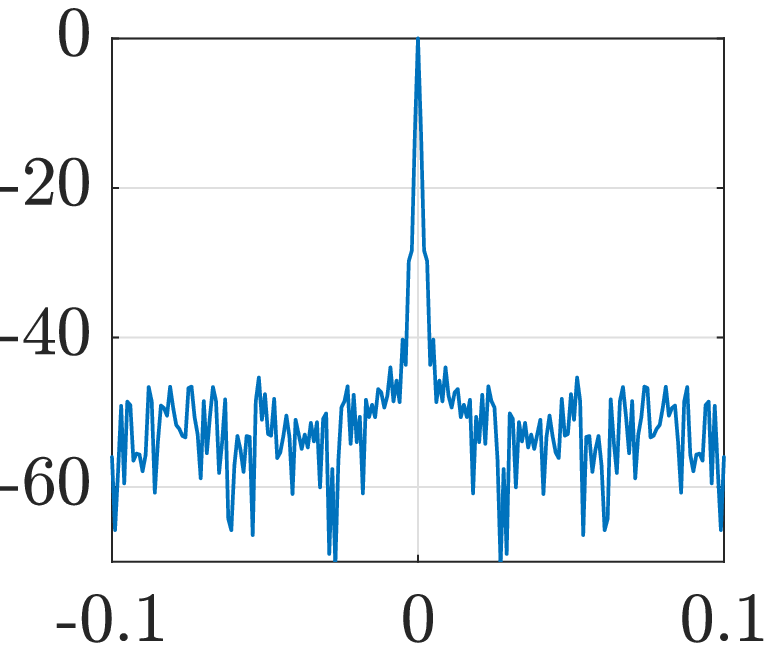}}
			\def\stackalignment{r}
			\topinset{\little}{\big}{15pt}{24pt}}}
          \mbox{\subfloat[]{\label{subfig10:d} 	\def\big{\includegraphics[width=0.48\textwidth,height=6.25cm]{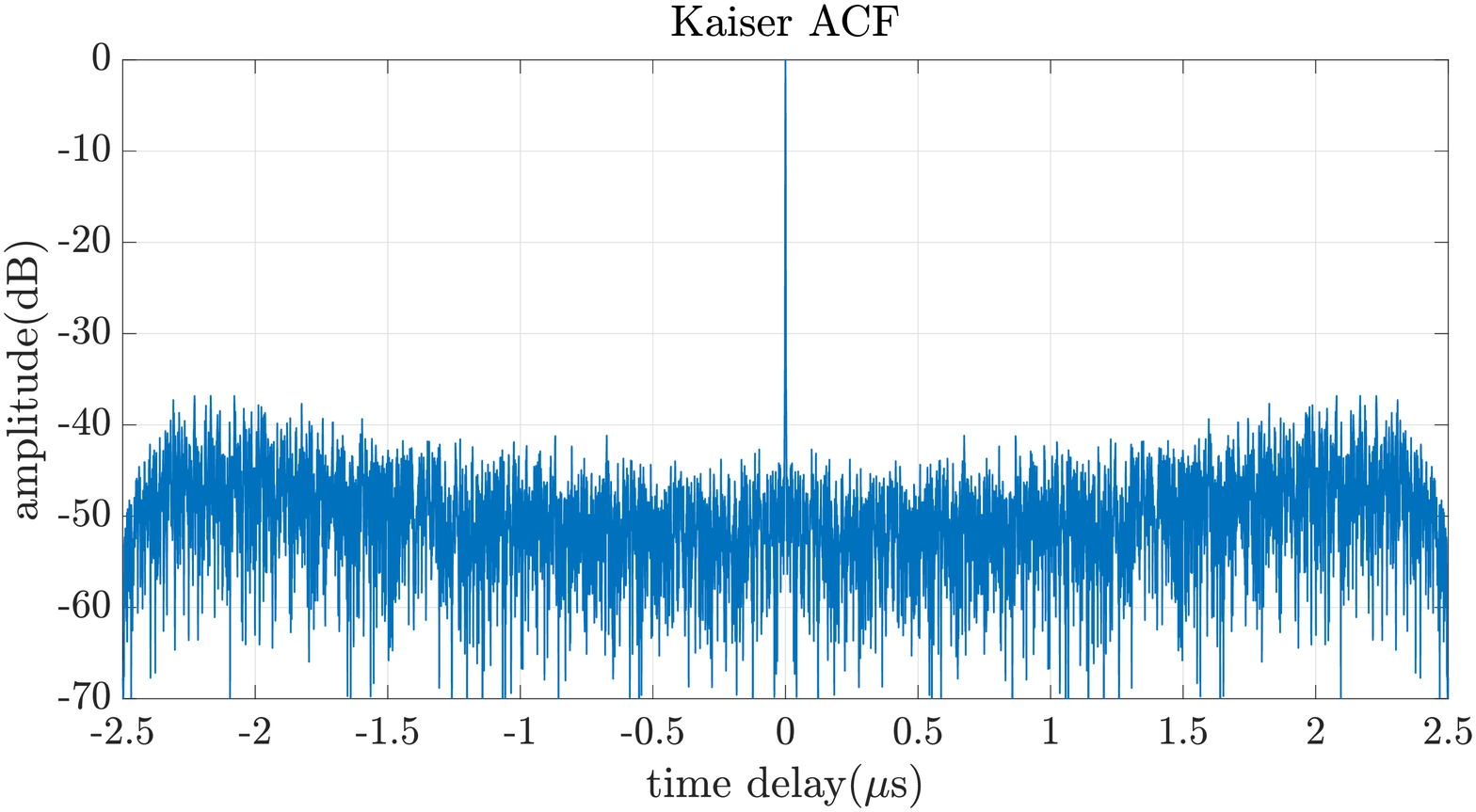}}
			\def\little{\includegraphics[height=2.4cm]{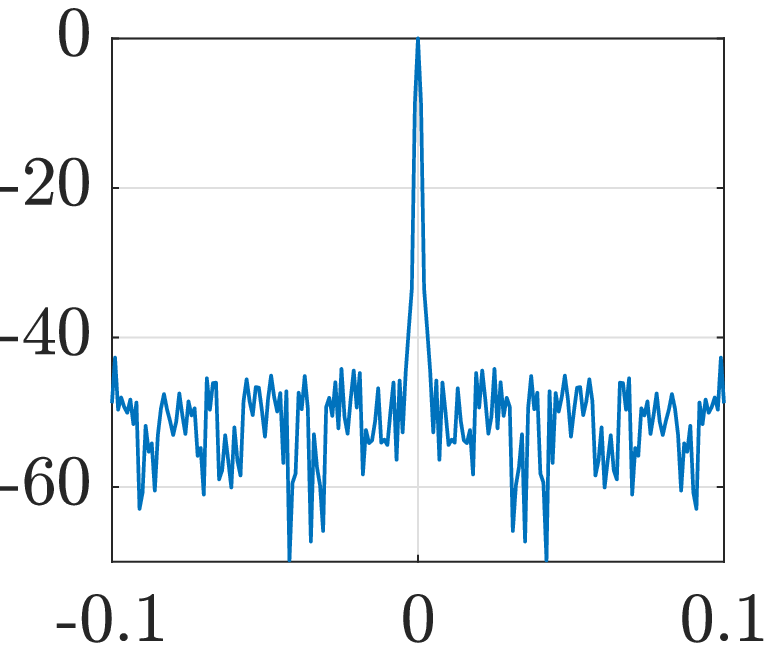}}
			\def\stackalignment{r}
			\topinset{\little}{\big}{15pt}{24pt}}}
	\caption{The autocorrelation functions of the designed
signals using the proposed method for the six windows of Raised-Cosine, Taylor, Chebyshev, Gaussian, Poisson, and Kaiser.}
	\label{fig:fig1}
\end{figure*}

\begin{table*}[!h]
	\centering
	\caption{Selected Windows.}
\resizebox{\textwidth}{!}{\begin{tabular}{c c c c}
                     \toprule[1.5pt]\\
		\bfseries Windows & \bfseries Formula & \bfseries Group Time Delay Function & \bfseries Constant Parameters\\ \\
                     \toprule[1.5pt]\\
		Raised-Cosine & $w\left( n \right) = k + \left( {1 - k} \right){\rm{co}}{{\rm{s}}^2}\left( {\cfrac{{\pi n}}{{M - 1}}} \right),\;\;\;\;\;\left| n \right| \le \cfrac{{M - 1}}{2}$ & $ {\cfrac{Tf}{B} + \cfrac{T}{{2\pi }}\left( {\cfrac{{1 - k}}{{1 + k}}} \right)\sin \left( {\cfrac{{2\pi f}}{B}} \right)} $  & $k = 0.17$\\ \\
		\hline\\
		Taylor & $\begin{array}{l}
w\left( n \right) = 1 + \displaystyle\sum \limits_{m = 1}^{\bar n - 1} {F_m}{\rm{cos}}\left( {\cfrac{{2\pi mn}}{{M - 1}}} \right),\;\;\;\;\;\left| n \right| \le \cfrac{{M - 1}}{2}\\
{F_m} = F\left( {m,\bar n,\eta } \right)
\end{array}$ & ${\cfrac{Tf}{B} + \cfrac{T}{{2\pi }} \displaystyle\sum \limits_{m = 1}^{\bar n - 1} \cfrac{{{F_m}}}{m}{\rm{sin}}\left( {\cfrac{{2\pi mf}}{B}} \right)} $ & $\begin{array}{l} 
		\eta  = 88.5\;dB\\
		\bar n = 2
		\end{array}$ \\ \\
		\hline \\
		Chebyshev & $\begin{array}{l}
W\left( m \right) = \cfrac{{\cos \left\{ {M{\rm{co}}{{\rm{s}}^{ - 1}}\left[ {\beta \cos \left( {\frac{{\pi m}}{M}} \right)} \right]} \right\}}}{{\cosh \left[ {M{\rm{cos}}{{\rm{h}}^{ - 1}}\left( \beta  \right)} \right]}}\;\;,\;\;\;\;m = 0,1,2, \ldots ,M - 1\\
\beta  = \cosh \left[ {\frac{1}{M}{\rm{cos}}{{\rm{h}}^{ - 1}}\left( {{{10}^\alpha }} \right)} \right]\\
w\left( n \right) = \cfrac{1}{N} \displaystyle\sum \limits_{m = 0}^{M - 1} W\left( m \right).{\rm{exp}}\left( {\cfrac{{j2\pi mn}}{M}} \right),\;\;\;\;\;\;\left| n \right| \le \cfrac{{M - 1}}{2}
\end{array}$ & Calculated Numerically & $\alpha  = 2$ \\ \\
		\hline\\
                      Gaussian& $w\left( n \right) = \exp \left( { - k{{\left( {\cfrac{n}{{2\left( {M - 1} \right)}}} \right)}^2}} \right),\;\;\;\;\;\;\left| n \right| \le \cfrac{{M - 1}}{2}$ & $\cfrac{T}{{2\;{\rm{erf}}\left( {\sqrt k /4} \right)}}\ {\rm{erf}}\left( {\cfrac{{f\sqrt k }}{{2B}}} \right)\;$& $k = 35.51$\\ \\
		\hline\\
                      Poisson& $w\left( n \right) = \exp \left( { - k\cfrac{{\left| n \right|}}{{\left( {M - 1} \right)}}} \right),\;\;\;\;\;\;\;\;\left| n \right| \le \cfrac{{M - 1}}{2}$ & $\cfrac{{T{\rm{sgn}}\left( f \right)}}{{2\left( {1 - \exp \left( { \cfrac{-k}{2}} \right)} \right)}}
\left( {1 - \exp \left( {\cfrac{{ - k\left| f \right|}}{B}} \right)} \right)$ & $k = 2.5$\\ \\
		\hline\\
		Kaiser & $w\left( n \right) \buildrel \Delta \over = \left\{ {\begin{array}{*{20}{c}}
{\cfrac{{{I_0}\left( {\pi \alpha \sqrt {1 - {{\left( {\frac{n}{{M/2}}} \right)}^2}} } \right)}}{{{I_0}\left( {\pi \alpha } \right)}}\;\;\;\;\;}&{\left| n \right| \le \cfrac{{M - 1}}{2}}\\
0&{elsewhere}
\end{array}} \right.,\;\beta  \buildrel \Delta \over = \pi \alpha $ & Calculated Numerically & $\beta  = 4.5$ \\ \\
                     \toprule[1.5pt]
	\end{tabular} }
\end{table*}

\begin{figure*}[!ht]
	\centering
           \includegraphics[width=\linewidth,height=10.2cm]{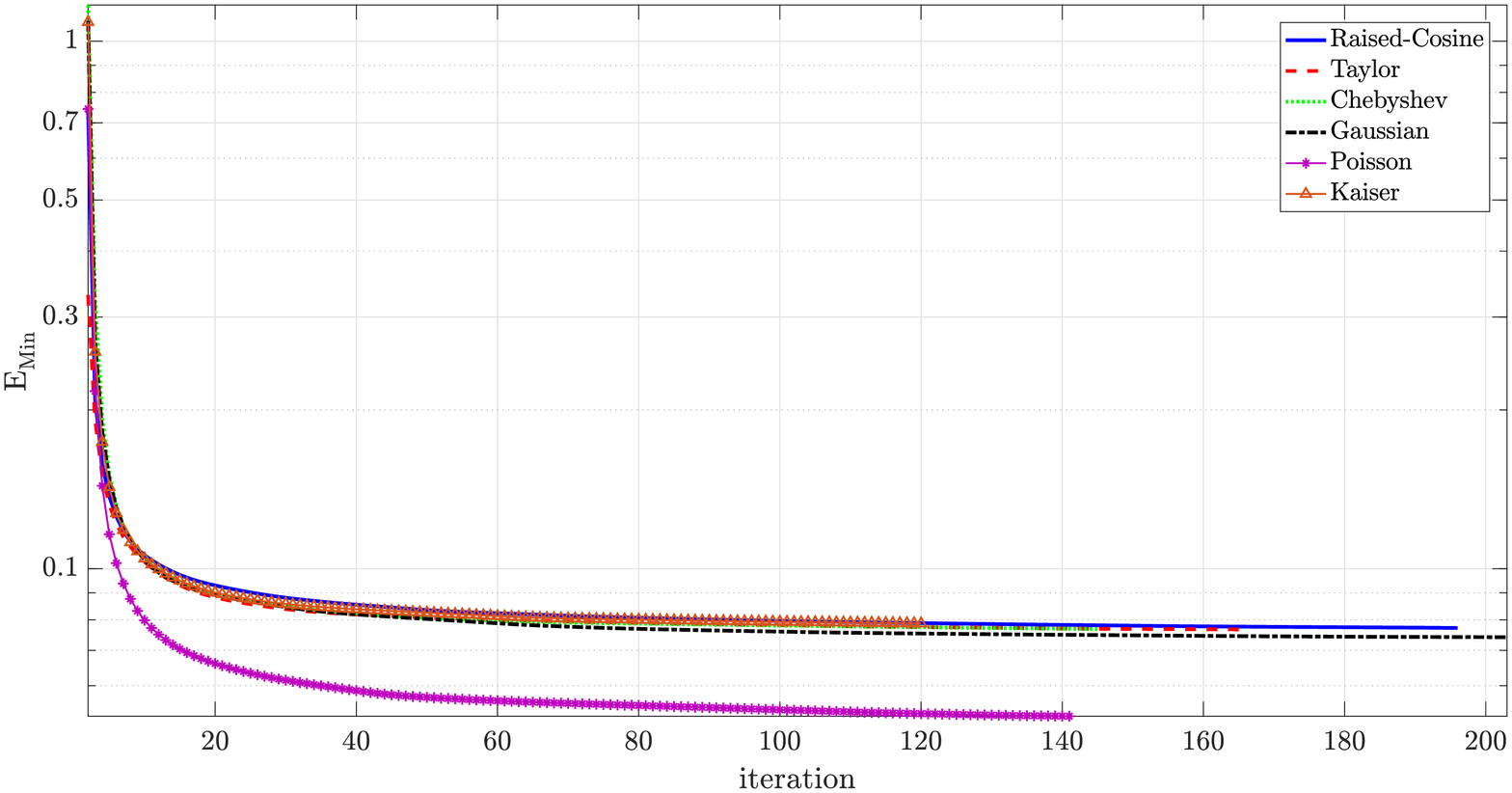}
	\caption{Minimum error with respect to iteration for the six windows of Raised-Cosine, Taylor, Chebyshev, Gaussian, Poisson, and Kaiser.}
	\label{fig:fig11}
\end{figure*}

Table 2 compares the results obtained from the stationary phase method and the proposed method for PSL of the autocorrelation function. The results indicate that the average of PSL reduction is about 5 dB revealing the maximum PSL reduction associated with the Poisson window with -17.28 dB decrement. The minimum error in Fig. 3 is calculated according to (28). As mentioned in section 3, the minimum error of the proposed method has a decreasing trend. First, this error has a significant value, but it begins to decrease and the trend of its change is almost constant in high iterations.

\begin{table}[!t]
\centering
\caption{Comparison of PSL for the Stationary Phase Method (SPM) and Pro-\\posed Method (PM).}
\begin{tabular}{c c c c}
          \toprule[1.5pt]
           \multirow{2}{*}{\bfseries Windows}& \multicolumn{2}{c} {\bfseries PSL (dB)} & \multirow{2}{*}{\bfseries Improvement (dB)}              
          \\
          \cmidrule(r){2-3}
	 & \bfseries  SPM& \bfseries PM & \\
          \toprule[1.5pt] \\
	Raised-Cosine & $-33.34$ & $-37.89$ & $-4.55$ \\ \\
	\hline \\
	Taylor & $-33.34$ & $-37.73$ & $-4.39$ \\ \\
	\hline \\
	Chebyshev & $-31.77$ & $-37.37$ & $-5.60$ \\ \\
	\hline \\
           Gaussian & $-32.38$ & $-37.67$ & $-5.29$ \\ \\
	\hline \\
           Poisson & $-20.39$ & $-37.67$ & $-17.28$ \\ \\
	\hline \\
	Kaiser & $-30.98$ & $-36.82$ & $-5.84$ \\ \\
          \bottomrule[1.5pt]
\end{tabular}
\end{table}

\section{Conclusion}
In the proposed method, a NLFM signal by solving a constrained optimization problem using Lagrangian method is obtained. By this iterative method, the PSL of the autocorrelation function reduced about 5 dB compared with the stationary phase method. PSL reduction for the Poisson window compared to other windows is significant. Using mathematical analysis, we showed that the minimum error of the proposed method has a decreasing trend and this guarantees the convergence of proposed method. The results of the minimum error of six selected windows also reveal the validity of this statement.

\end{document}